\documentclass{article}
\usepackage{spconf,amsmath,graphicx}
\usepackage{amsmath}
\usepackage{amssymb}
\usepackage{amsthm}
\usepackage{amsfonts}

\usepackage{amsmath,amsfonts,bm}
\DeclareMathOperator*{\st}{ \quad \textnormal{ s.t. }}%
\def\eqref#1{equation~\ref{#1}}

\def\1{\bm{1}}

\DeclareMathAlphabet{\mathsfit}{\encodingdefault}{\sfdefault}{m}{sl}
\SetMathAlphabet{\mathsfit}{bold}{\encodingdefault}{\sfdefault}{bx}{n}

\newcommand{\R}{\mathbb{R}}

\DeclareMathOperator*{\argmin}{arg\,min}

\usepackage{placeins}
\usepackage{float}

\usepackage{xcolor}
\usepackage[usestackEOL]{stackengine}

\def\R{{\mathbb R}}

\title{Strong Data Augmentation Sanitizes Poisoning and Backdoor Attacks Without an Accuracy Tradeoff}
\name{\Longstack{Eitan Borgnia* \, Valeriia Cherepanova* \, Liam Fowl* \, Amin Ghiasi* \, Jonas Geiping $\dagger$ \\
      Micah Goldblum* \, Tom Goldstein* \, Arjun Gupta*}
      \thanks{Authors ordered alphabetically}}

\address{*University of Maryland, College Park \qquad $\dagger$ University of Siegen}

\begin{document}

\maketitle

\begin{abstract}

\sloppy

Data poisoning and backdoor attacks manipulate victim models by maliciously modifying training data. In light of this growing threat, a recent survey of industry professionals revealed heightened fear in the private sector regarding data poisoning.  Many previous defenses against poisoning either fail in the face of increasingly strong attacks, or they significantly degrade performance. However, we find that strong data augmentations, such as mixup and CutMix, can significantly diminish the threat of poisoning and backdoor attacks without trading off performance.  We further verify the effectiveness of this simple defense against adaptive poisoning methods, and we compare to baselines including the popular differentially private SGD (DP-SGD) defense.  In the context of backdoors, CutMix greatly mitigates the attack while simultaneously increasing validation accuracy by 9\%.
\end{abstract}
\begin{keywords}
Data Poisoning, Backdoor Attacks, Adversarial Attacks, Differential Privacy, Data Augmentation
\end{keywords}
\section{Introduction}
\label{sec:intro}

Machine learning models have demonstrated tremendous success in many domains including mobile image processing \cite{Schwartz_2019}, targeted ad placement \cite{fang_poisoning_2018}, and security services \cite{lovisotto_biometric_2019}. The growing availability of vast datasets has aided in this recent success. Practitioners often rely upon data scraped from the web or sourced from a third party \cite{taigman_deepface:_2014,papernot_marauders_2018}, where the security of such data can be compromised by malicious actors. \textit{Data Poisoning} attacks pose a specific threat in which an attacker modifies a victim's training data to achieve goals such as targeted misclassification or performance degradation. Basic data poisoning schemes implement backdoor triggers in training data, whereas recent works have also demonstrated that data poisoning schemes can successfully attack deep learning models trained on industrial-scale datasets \cite{huang2020metapoison, geiping2020witches} without perceptible modifications. The seriousness of these threats is acknowledged by industrial practitioners, who recently ranked poisoning as the most worrisome threat to their interests, in a recent study \cite{kumar_adversarial_2020}. Furthermore, defenses designed for older, less powerful poisoning strategies work by filtering out poisons based on feature anomalies \cite{rubinstein_antidote:_2009} but fail when models are trained from scratch on poisoned data \cite{peri2019deep, geiping2020witches}. Currently, the only method to prevent state-of-the-art targeted poisoning relies upon differentially private SGD (DP-SGD) and leads to a significant drop in validation accuracy \cite{geiping2020witches, schwarzschild2020just,hong_effectiveness_2020}.  

On the other hand, data augmentation has been a boon to practitioners, aiding in state-of-the-art performance on a variety of tasks \cite{zhang2017mixup, gong2020maxup}. Data augmentation can be used in many regimes, including settings where data is sparse, to improve generalization \cite{ni2020data}. Simple augmentations include random crops or horizontal flips. Recently, more sophisticated augmentation schemes have emerged that improve model performance: {\em mixup} takes pairwise convex combinations of randomly sampled training data and uses the corresponding convex combinations of labels. Not only does this prevent memorization of corrupt labels and provide robustness to adversarial examples, but it has also been shown to improve generalization \cite{zhang2017mixup}. Another augmentation technique, cutout, randomly erases patches of training data \cite{devries2017improved}, whereas CutMix instead combines pairwise randomly sampled training data by taking random patches from one image and overlaying these patches onto other images \cite{yun2019cutmix}. The labels are then mixed proportionally to the area of these patches. CutMix enhances model robustness, achieves better test accuracy, and improves localization ability by encouraging the network to correctly classify images from a partial view. Finally, MaxUp applies a set of data augmentation techniques (basic or complex) to the training data and chooses the augmentation method and parameters that achieve the worst model performance of all the techniques \cite{gong2020maxup}. By training against the most ``difficult" data augmentation, MaxUp is able to improve generalization and, in some cases, adversarial robustness. 

We investigate the effects of multiple augmentation strategies on data poisoning attacks. We find that these modern data augmentation strategies are often more effective than previous more cumbersome defenses against poisoning while also not sacrificing significant natural validation accuracy. 
Our data augmentation defense is additionally convenient for practitioners as it involves only a small and easy-to-implement change to standard training pipelines.

We empirically analyze this defense both in the setting of a simple backdoor trigger attack and in the setting of a modern imperceptible targeted data poisoning attack \cite{geiping2020witches}.

\section{Threat Model}
There are many different flavors of poisoning attacks. In this paper, we focus on two attacks from both sides of the spectrum: A simple and robust backdoor trigger attack and a modern poisoning attack that is \textit{targeted} and also \textit{clean-label}. Backdoor trigger attacks insert a specific trigger pattern (usually a small patch, but sometimes an additive perturbation or non-rectangular symbol) into training data.  If this pattern is then added to images at test time, the network will misclassify the test image, assigning the label that was placed on the training images poisoned at train time. In contrast, \textit{targeted} attacks are those in which the attacker wishes to modify the victim model to specifically misclassify a set of target images at inference time. Such attacks are often {\em clean-label}, meaning the modified training images retain their semantic content and are labeled correctly. Such attacks are optimization-based, finding the most effective perturbation of training data using gradient descent. This can make the attack especially hard to detect for sanitization-based defenses because it does not significantly degrade clean accuracy \cite{huang2020metapoison, geiping2020witches}. We focus on both backdoor and clean-label attacks because of their renewed and recent interest in the community and because they cover the poisoning literature from two sides.

Any poisoning threat model can be formally described as a bilevel problem. 
Let $F(x, \theta)$ be a neural network taking inputs $x \in \R^n$ with parameters $\theta \in \mathbb{R}^p$. The attacker is allowed to modify $P$ samples out of $N$ total samples (where $P \ll N$) by adding perturbation $\Delta_{i}$ to the $i^{th}$ training image. The perturbation is constrained in the $\ell_0$ norm for the patch-based/backdoor trigger attack, or the  $\ell_\infty$-norm in the case of optimization-based attacks we consider. 

Attackers wish to find $\Delta$ so that a set of $T$ target samples $(x^t_i, y^t_i)_{i=1}^T$ are classified with the new, incorrect, adversarial labels $y^\text{adv}_i$ after training by minimizing loss function $\mathcal{L}$:
 \begin{gather}\label{eq:bilevel}
     \min_{\Delta \in \mathcal{C}} \sum_{i=1}^T \mathcal{L} \left( F(x^t_i, \theta(\Delta)), y^\text{adv}_i\right)\\
  \st \theta(\Delta) \in \argmin_\theta \frac{1}{N} \sum_{i=1}^N \mathcal{L}(F(x_i + \Delta_i, \theta), y_i).
 \end{gather}
In this framework, we can understand \textit{backdoor} attacks as choosing the optimal $\Delta$ directly based on a given rule (here via patch insertion onto training images with the target label $y_i^\text{adv}$), whereas optimization-based methods such as \cite{geiping2020witches} optimize some approximation of the (intractable) full bilevel optimization problem. Witches' Brew \cite{geiping2020witches} approximately optimizes $\Delta$ by modifying training data so the gradient of the training objective is aligned with the gradient of the adversarial loss $\mathcal{L} \left( F(x^t_i, \theta(\Delta)), y^\text{adv}_i\right)$, using optimization methods based on adversarial literature \cite{madry_towards_2017,chiang2020witchcraft, abdelkader2020headless}. 

Both backdoor attacks and targeted data poisoning attacks rely upon the expressiveness of modern deep networks trained from scratch in order to ``gerrymander" the network's decision boundary around specific target images \cite{geiping2020witches} \textemdash $\text{}$ they behave normally on validation data and the chosen target is often made into a class outlier that is surgically labeled in an adversarial way using an erratic decision boundary. As a result, one can assume changes in the training procedure, such as strong data augmentation, could have a significant impact on the success of poisoning methods by imposing regularity on the decision boundary and preventing target images from being gerrymandered into the wrong class. Note the sensitivity of poisoning to changes in training has been confirmed by defenses involving gradient clipping/noising, which is the only defense shown in \cite{geiping2020witches,hong_effectiveness_2020} to degrade poisoning success. However, this defense ultimately proves impractical because of the decreased natural accuracy that comes with robustness to poisoning. Thus, we aim to bridge this gap and develop small changes in training via data augmentation that defend against data poisoning without impeding normal training. 

\section{Method}


Mixup can be interpreted as a method for convexifying class regions in the input space \cite{zhang_mixup:_2017}.  By enforcing that convex combinations of training points are assigned convex combinations of the labels, this augmentation method regularizes class boundaries, and removes small non-convex regions.  In particular, we are motivated by the idea of using mixup to promote the removal of small ``gerrymandered'' regions in input space in which a target/poisoned data instance is assigned an adversarial label while being surrounded by (non-poisoned) instances with different labels. 

In our experiments, we use a generalization of the mixup process from \cite{zhang_mixup:_2017} for mixture width $k$. Instead of a Beta distribution, convex coefficients are drawn from a Dirichlet distribution $\text{Dir}[\alpha, \dots, \alpha]$ of order $k$ with interpolation parameter $\alpha = 1.$

\begin{table*}[h]
\begin{tabular}{c|c c c c c }
    \textbf{} &  Poison Success (100\%) & Validation Accuracy (100\%) & Poison Success (10\%) & Validation Accuracy (10\%) \\
    \hline 
    Baseline & 100\% & 85\% & 57\% & 94\% \\
    mixup &    100\% & 85\% & 42\% & 95\% \\
    CutMix &   36\%  & 94\% & 23\% & 95\% \\
\end{tabular}
\caption{Poison success and (clean) validation accuracy in the from-scratch setting against backdoor attacks. The first two columns correspond to poisoning all images in the target class, the last two columns correspond to poisoning 10\% of images in the target class. All values are averaged over 4 runs.}
\label{tab:backdoor}
\end{table*}

\sloppy
We implement CutMix \cite{devries_improved_2017} as follows: Let $\mathcal{D} = \{(x_i, y_i)\}_{i=1}^n$ be the training dataset with $x_i \in \mathbb{R}^{w \times h \times c}$. Let $(x_i,y_i)$ and $(x_j, y_j)$ be two randomly sampled feature-target pairs. We randomly generate a box $M \in \{0,1\}^{w \times h}$ that indicates the pixels to be cut/pasted. All values of $M$ are defined to be one except in a box centered at $(r_x,r_y)$ where they are zero. To obtain the box location, we randomly sample the center $r_x \sim \text{Unif}(0,w)$ and $r_y \sim \text{Unif}(0,h)$ for the first and second axes respectively. As in mixup, we use a coefficient $\lambda \sim \text{Dir}[1,1]$ to determine the relative contribution from each of the two randomly sampled data points. i.e. $r_w = w \sqrt{1-\lambda}$ and $r_h = h\sqrt{1-\lambda}$ give the width and height for the patch of zeros in $M$. The augmented image for CutMix becomes $\Tilde{x} = M \bigodot x_i + (1-M) \bigodot x_j$ with label $\Tilde{y} = \lambda y_i + (1-\lambda) y_j$ obtained by mixing initial labels proportionally to the size of $M$.  The binary operation $\bigodot$ represents element-wise multiplication. 

Cutout is similar, but the patch remains black. The cutout data point is given by $\Tilde{x} = M \bigodot x_i$ with label $y_i$.

The procedure for MaxUp is taken from \cite{gong2020maxup}: For each data point $x_i$ in the original training set $\mathcal{D} = \{(x_i, y_i)\}_{i=1}^n$, a set $\{\Tilde{x}_{i,j}\}_{j=1}^{m}$ of augmented data points are produced. Learning is defined according to a modified ERM,
$$\min_{\theta}\mathbb{E}_{x \sim \mathcal{D}}[\max_{j}\mathcal{L}(\Tilde{x}_{i,j}, \theta)].$$


\begin{figure}[h]
    \centering
    \includegraphics[scale=.35]{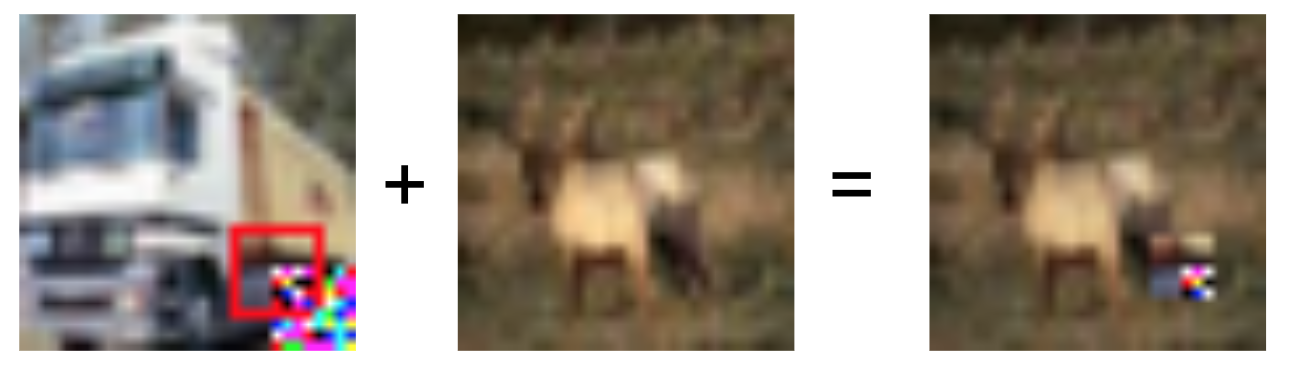}
    \caption{An illustration of CutMix with poisoned data (truck) and non-poisoned data (deer).  
    CutMix chops up the patch, lessening its visual impact in training data.}
    \label{fig:mixup_pic}
\end{figure}

\section{Experiments}

\begin{table*}[t]
\centering
\begin{tabular}{c|c c c }
    \textbf{} & Basic Attack (\%) & Adaptive Attack (\%)  & Validation Accuracy (\%)\\
    \hline 
    Baseline & 90.00 & 90.00  & 92.08 \\
    \hline
    DP-SGD, $0.01$  & 77.00  & 86.00  & 91.33 \\
    DP-SGD, $0.05$  & 1.00   & 40.00  & 81.62 \\
    \hline
    mixup &  45.00 & 72.00  & 91.50\\
    mixup (4-way) &  5.00  & 55.00  & 87.45\\
    CutMix & 75.00  & 60.00   & 91.62\\
    cutout & 60.00  & 81.25 &  91.64 \\
    MaxUp-cutout & 5.00 & 20.00   & 86.05\\
\end{tabular}
\caption{Poison success and (clean) validation error results for the from-scratch setting against Witches' Brew. 
All values are averaged over 20 runs.  Lower values in the first two columns are better. Higher numbers in the last column are better.}
\label{tab:witchesbrew}
\end{table*}

\subsection{Defending Against Backdoor Trigger Attacks}

We first demonstrate the effectiveness of data augmentation at mitigating backdoor attacks while at the same time increasing test accuracy.  To establish baselines for backdoor attacks, we train a ResNet-18 \cite{he_deep_2015} on the CIFAR-10 dataset consisting of $50,000$ images in $10$ balanced classes \cite{krizhevsky2009learning}. We insert triggers into the training set by adding $4 \times 4$ patches to training images in a randomly selected \emph{target class}. 
We then evaluate on patched images from a new \emph{base class} to see if the patch triggers images to be misclassified into the target class. We perform this experiment in two different settings: when all training images are patched in the base class and only 10\% of training images are patched in this class. These two scenarios reflect varying access an adversary might have to their victim's training data.

We report the success of this attack at causing base images to be misclassified with the target label in Table \ref{tab:backdoor}. Although mixup data augmentation does not defend against the backdoor attack, CutMix dramatically reduces the success rate of the poison attack from $100\%$ to $36\%$ while simultaneously increasing validation accuracy by $9\%$.  One explanation for the ineffectiveness of mixup in this domain is that, under this strategy, the base class can still be associated with the patch. On the other hand, CutMix randomly replaces parts of the images and therefore may cut patches apart.  Moreover, CutMix improves clean validation accuracy since the network learns relevant features from the target class rather than simply relying on the patch.

\begin{figure}[h]
    \centering
    \includegraphics[scale=.35]{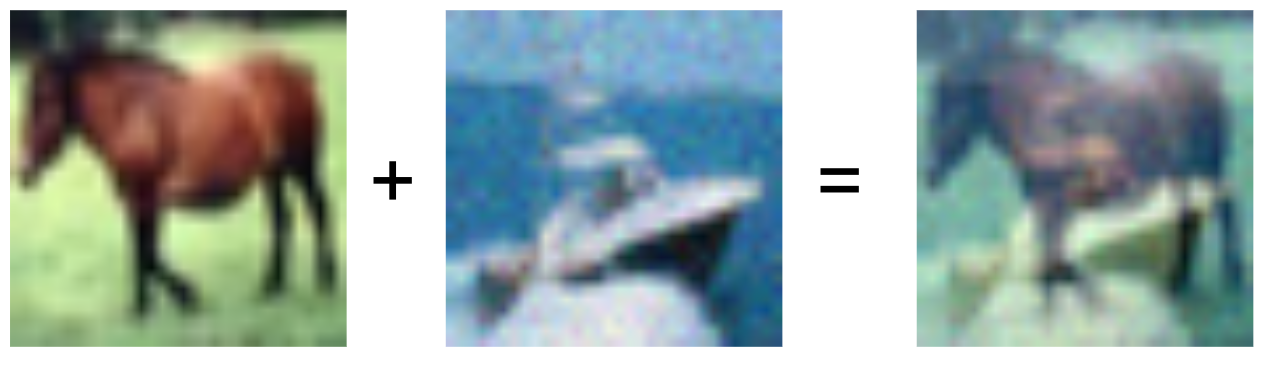}
    \caption{Under mixup, perturbed images are averaged with clean images.}
    \label{fig:mixup_pic2}
    \vspace{-0.5cm}
\end{figure}

\subsection{Defending Against Targeted Poisoning Attacks}
\begin{figure*}[h]
    \centering
    \vspace{-0.1cm}
    \includegraphics[trim=0 0 0 35mm, clip, width=0.4\textwidth]{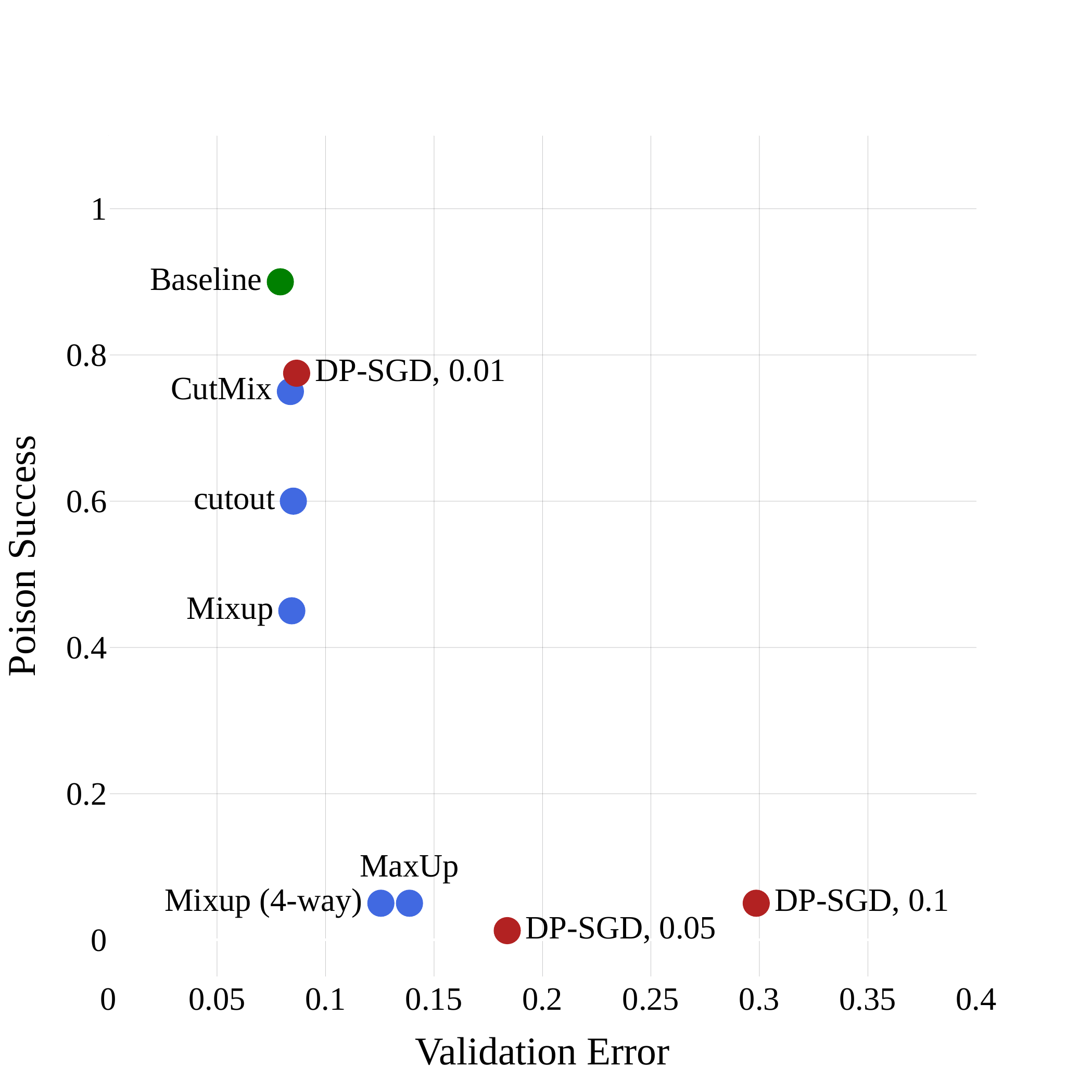}
    \includegraphics[trim=0 0 0 35mm, clip, width=0.4\textwidth]{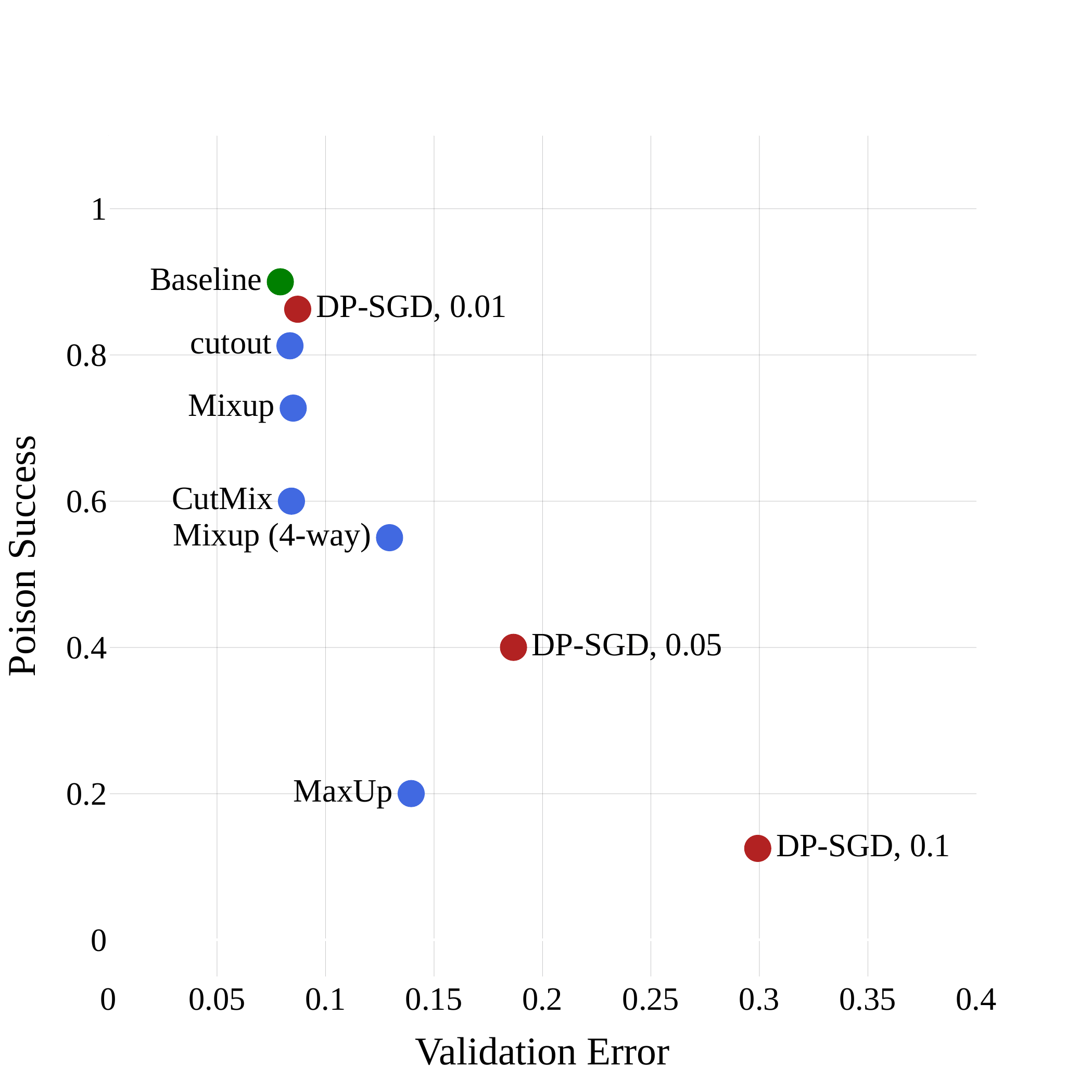}
    \caption{Poison success vs. (clean) validation accuracy for non-adaptive attacks (left) and adaptive attacks (right).}
    \label{fig:scatterthing}
    \vspace{-0.35cm}
\end{figure*}

We now evaluate data augmentation as a defense against targeted poisoning attacks.  To this end, we use the state-of-the-art method, Witches' Brew~\cite{geiping2020witches}, as well as an ``adaptive attack'' version in which attacks are generated on networks trained with the same data augmentations as the victim.  We train a ResNet-18 on CIFAR-10, and we consider a threat model with $\ell_\infty$ bound  $16/255$ and a budget of $1\%$. For all experiments, we consider a baseline model trained with random horizontal flips and random crops. We then compare modern data augmentation against differential-privacy based defenses with a special focus on practical clean validation accuracy, that is ``normal'' usability of the model. We conduct and average success over 20 runs, where each run consists of a randomly chosen target image and a randomly chosen adversarial label.  In each run, a new model is trained from scratch on the poisoned training set and evaluated on the target image.  The lower the attack's success rate, the more effective the defense.

Table~\ref{tab:witchesbrew} shows that we can defend using differentially private SGD (DP-SGD) \cite{abadi_deep_2016} with sufficient amounts of Gaussian noise added to all training gradients \textemdash $\text{}$ but this comes at a tremendous costs in validation accuracy. 
In this setting of an optimization-based attack, we also have to contend with another factor: \textit{adaptive attacks}. An adaptive attack threat model assumes that the attacker is aware of the defense and can optimize their attack w.r.t to this defense. This attack is highlighted in the second column of Table~\ref{tab:witchesbrew}. In this work we adapt WitchesBrew to advanced data augmentation by incorporating the data augmentation into the training phase of the clean model used in the attack. This way poisoned data is created based on a model trained to be invariant to these augmentations. We adapt differential privacy as in \cite{geiping2020witches} by a straight-through estimate, finding it to be mitigated especially well by an adaptive attack.

In contrast to the loss of validation accuracy incurred by DP-SGD defenses, the defenses via data augmentation (lower rows in Table~\ref{tab:witchesbrew}, blue dots in Figure~\ref{fig:scatterthing}) lose almost no validation accuracy. Nonetheless, they reduce poison success by, for example, up to $60\%$ in the case of MaxUp based on four cutouts. This improvement supports the conclusion of \cite{gong2020maxup} that MaxUp implicitly imposes a gradient penalty, thus smoothing decision boundaries.  Also notable is that for this case of optimization-based poisoning attacks, mixup seems to be the optimal data augmentation among those tried which do not degrade validation accuracy. We conjecture that this is due to the implicit linearity enforced between data points by mixup optimization, which makes it harder for the poisoning scheme to successfully generate outliers. We can even further increase the defensive capabilities of mixup by considering a four-fold mixture of images instead of the two-fold mixtures discussed in \cite{zhang_mixup:_2017}. This defense is even stronger than vanilla mixup in both cases, but the four-fold mixtures of images leads to data modifications so stark that they negatively influence validation accuracy. 

In comparison to the backdoor trigger attacks, CutMix is less effective against the non-adaptive attack. However, we also find that for the adaptive setting, CutMix improves upon mixup.  This is likely because the entire image perturbation appears in the mixed image when mixup is used, enabling reliable poisoning.  In contrast, an adaptive attack on CutMix cannot \textit{a priori} know the location of the cut patch, hence the attack is impeded even if it is known that CutMix is used. 
\section{Conclusions}
Data poisoning attacks are increasingly threatening to machine learning practitioners. 
By and large, defenses have not kept pace with rapidly improved attacks, and these defenses can be impractical to implement. We demonstrate that modern data augmentation schemes can adjust training behavior of neural networks to mitigate from-scratch poisoning while maintaining natural accuracy for both backdoor triggers and optimization-based attacks.  We think these results suggest the possibility of specially-designed augmentations for poison defense, and we think this may be a fruitful direction for future research.

\bibliographystyle{IEEEbib}
\bibliography{refs, zotero_library}

\begin{thebibliography}{10}

\bibitem{Schwartz_2019}
Eli Schwartz, Raja Giryes, and Alex~M. Bronstein,
\newblock ``Deepisp: Toward learning an end-to-end image processing pipeline,''
\newblock {\em IEEE Transactions on Image Processing}, vol. 28, no. 2, pp.
  912–923, Feb 2019.

\bibitem{fang_poisoning_2018}
Minghong Fang, Guolei Yang, Neil~Zhenqiang Gong, and Jia Liu,
\newblock ``Poisoning {{Attacks}} to {{Graph}}-{{Based Recommender Systems}},''
\newblock in {\em Proceedings of the 34th {{Annual Computer Security
  Applications Conference}}}, {San Juan, PR, USA}, Dec. 2018, {{ACSAC}} '18,
  pp. 381--392, {Association for Computing Machinery}.

\bibitem{lovisotto_biometric_2019}
Giulio Lovisotto, Simon Eberz, and Ivan Martinovic,
\newblock ``Biometric {{Backdoors}}: {{A Poisoning Attack Against Unsupervised
  Template Updating}},''
\newblock {\em ArXiv190509162 Cs}, May 2019.

\bibitem{taigman_deepface:_2014}
Yaniv Taigman, Ming Yang, Marc'Aurelio Ranzato, and Lior Wolf,
\newblock ``{{DeepFace}}: {{Closing}} the {{Gap}} to {{Human}}-{{Level
  Performance}} in {{Face Verification}},''
\newblock in {\em 2014 {{IEEE Conference}} on {{Computer Vision}} and {{Pattern
  Recognition}}}, {Columbus, OH, USA}, June 2014, pp. 1701--1708, {IEEE}.

\bibitem{papernot_marauders_2018}
Nicolas Papernot,
\newblock ``A {{Marauder}}'s {{Map}} of {{Security}} and {{Privacy}} in
  {{Machine Learning}},''
\newblock {\em ArXiv181101134 Cs}, Nov. 2018.

\bibitem{huang2020metapoison}
W~Ronny Huang, Jonas Geiping, Liam Fowl, Gavin Taylor, and Tom Goldstein,
\newblock ``Metapoison: Practical general-purpose clean-label data poisoning,''
\newblock {\em arXiv preprint arXiv:2004.00225}, 2020.

\bibitem{geiping2020witches}
Jonas Geiping, Liam Fowl, W~Ronny Huang, Wojciech Czaja, Gavin Taylor, Michael
  Moeller, and Tom Goldstein,
\newblock ``Witches' brew: Industrial scale data poisoning via gradient
  matching,''
\newblock {\em arXiv preprint arXiv:2009.02276}, 2020.

\bibitem{kumar_adversarial_2020}
Ram Shankar~Siva Kumar, Magnus Nyström, John Lambert, Andrew Marshall, Mario
  Goertzel, Andi Comissoneru, Matt Swann, and Sharon Xia,
\newblock ``Adversarial {{Machine Learning}} -- {{Industry Perspectives}},''
\newblock {\em ArXiv200205646 Cs Stat}, May 2020.

\bibitem{rubinstein_antidote:_2009}
Benjamin~I.P. Rubinstein, Blaine Nelson, Ling Huang, Anthony~D. Joseph,
  Shing-hon Lau, Satish Rao, Nina Taft, and J.~D. Tygar,
\newblock ``{{ANTIDOTE}}: {{Understanding}} and {{Defending Against Poisoning}}
  of {{Anomaly Detectors}},''
\newblock in {\em Proceedings of the 9th {{ACM SIGCOMM Conference}} on
  {{Internet Measurement}}}, {New York, NY, USA}, 2009, {{IMC}} '09, pp. 1--14,
  {ACM}.

\bibitem{peri2019deep}
Neehar Peri, Neal Gupta, W.~Ronny Huang, Liam Fowl, Chen Zhu, Soheil Feizi, Tom
  Goldstein, and John~P. Dickerson,
\newblock ``Deep k-nn defense against clean-label data poisoning attacks,''
  2019.

\bibitem{schwarzschild2020just}
Avi Schwarzschild, Micah Goldblum, Arjun Gupta, John~P Dickerson, and Tom
  Goldstein,
\newblock ``Just how toxic is data poisoning? a unified benchmark for backdoor
  and data poisoning attacks,''
\newblock {\em arXiv preprint arXiv:2006.12557}, 2020.

\bibitem{hong_effectiveness_2020}
Sanghyun Hong, Varun Chandrasekaran, Yiğitcan Kaya, Tudor Dumitraş, and
  Nicolas Papernot,
\newblock ``On the {{Effectiveness}} of {{Mitigating Data Poisoning Attacks}}
  with {{Gradient Shaping}},''
\newblock {\em ArXiv200211497 Cs}, Feb. 2020.

\bibitem{zhang2017mixup}
Hongyi Zhang, Moustapha Cisse, Yann~N Dauphin, and David Lopez-Paz,
\newblock ``mixup: Beyond empirical risk minimization,''
\newblock {\em arXiv preprint arXiv:1710.09412}, 2017.

\bibitem{gong2020maxup}
Chengyue Gong, Tongzheng Ren, Mao Ye, and Qiang Liu,
\newblock ``Maxup: A simple way to improve generalization of neural network
  training,''
\newblock {\em arXiv preprint arXiv:2002.09024}, 2020.

\bibitem{ni2020data}
Renkun Ni, Micah Goldblum, Amr Sharaf, Kezhi Kong, and Tom Goldstein,
\newblock ``Data augmentation for meta-learning,'' 2020.

\bibitem{devries2017improved}
Terrance DeVries and Graham~W Taylor,
\newblock ``Improved regularization of convolutional neural networks with
  cutout,''
\newblock {\em arXiv preprint arXiv:1708.04552}, 2017.

\bibitem{yun2019cutmix}
Sangdoo Yun, Dongyoon Han, Seong~Joon Oh, Sanghyuk Chun, Junsuk Choe, and
  Youngjoon Yoo,
\newblock ``Cutmix: Regularization strategy to train strong classifiers with
  localizable features,''
\newblock in {\em Proceedings of the IEEE International Conference on Computer
  Vision}, 2019, pp. 6023--6032.

\bibitem{madry_towards_2017}
Aleksander Madry, Aleksandar Makelov, Ludwig Schmidt, Dimitris Tsipras, and
  Adrian Vladu,
\newblock ``Towards {{Deep Learning Models Resistant}} to {{Adversarial
  Attacks}},''
\newblock {\em ArXiv170606083 Cs Stat}, June 2017.

\bibitem{chiang2020witchcraft}
Ping-Yeh Chiang, Jonas Geiping, Micah Goldblum, Tom Goldstein, Renkun Ni,
  Steven Reich, and Ali Shafahi,
\newblock ``Witchcraft: Efficient pgd attacks with random step size,''
\newblock in {\em ICASSP 2020-2020 IEEE International Conference on Acoustics,
  Speech and Signal Processing (ICASSP)}. IEEE, 2020, pp. 3747--3751.

\bibitem{abdelkader2020headless}
Ahmed Abdelkader, Michael~J Curry, Liam Fowl, Tom Goldstein, Avi Schwarzschild,
  Manli Shu, Christoph Studer, and Chen Zhu,
\newblock ``Headless horseman: Adversarial attacks on transfer learning
  models,''
\newblock in {\em ICASSP 2020-2020 IEEE International Conference on Acoustics,
  Speech and Signal Processing (ICASSP)}. IEEE, 2020, pp. 3087--3091.

\bibitem{zhang_mixup:_2017}
Hongyi Zhang, Moustapha Cisse, Yann~N. Dauphin, and David {Lopez-Paz},
\newblock ``Mixup: {{Beyond Empirical Risk Minimization}},''
\newblock {\em ArXiv171009412 Cs Stat}, Oct. 2017.

\bibitem{devries_improved_2017}
Terrance DeVries and Graham~W. Taylor,
\newblock ``Improved {{Regularization}} of {{Convolutional Neural Networks}}
  with {{Cutout}},''
\newblock {\em ArXiv170804552 Cs}, Aug. 2017.

\bibitem{he_deep_2015}
Kaiming He, Xiangyu Zhang, Shaoqing Ren, and Jian Sun,
\newblock ``Deep {{Residual Learning}} for {{Image Recognition}},''
\newblock {\em ArXiv151203385 Cs}, Dec. 2015.

\bibitem{krizhevsky2009learning}
Alex Krizhevsky, Geoffrey Hinton, et~al.,
\newblock ``Learning multiple layers of features from tiny images,''
\newblock 2009.

\bibitem{abadi_deep_2016}
Martin Abadi, Andy Chu, Ian Goodfellow, H.~Brendan McMahan, Ilya Mironov, Kunal
  Talwar, and Li~Zhang,
\newblock ``Deep {{Learning}} with {{Differential Privacy}},''
\newblock in {\em Proceedings of the 2016 {{ACM SIGSAC Conference}} on
  {{Computer}} and {{Communications Security}}}, {Vienna, Austria}, Oct. 2016,
  {{CCS}} '16, pp. 308--318, {Association for Computing Machinery}.

\end{thebibliography}

\end{document}